## *Magnetoresistance and spintronic anisotropy induced by spin excitations along molecular spin chains*

K. Katcko[1]*, E. Urbain[1]*, L. Kandpal[1], B. Chowrira[1,2], F. Schleicher[1], U. Halisdemir[1], F. Ngassam[1], D. Mertz[1], B. Leconte[1], N. Beyer[1], D. Spor[1], P. Panissod[1], A. Boulard[1], J. Arabski[1], C. Kieber[1], E. Sternitsky[1], V. Da Costa[1], M. Alouani[1], M. Hehn[3], F. Montaigne[3], A. Bahouka[4], W. Weber[1], E. Beaurepaire[1†], D. Lacour[3], S. Boukari[1], M. Bowen[1@]

[1] *Institut de Physique et Chimie des Matériaux de Strasbourg, UMR 7504 CNRS, Université de Strasbourg, 23 Rue du Lœss, BP 43, 67034 Strasbourg, France.*

[2] *Synchrotron SOLEIL, L'Orme des Merisiers, Saint-Aubin, BP 48, 91192 Gif-sur-Yvette, France*

[3] *Institut Jean Lamour UMR 7198 CNRS, Université de Lorraine, BP 70239, 54506 Vandœuvre les Nancy, France.*

[4] *IREPA LASER, Institut Carnot MICA, Parc d'innovation - Pole API, 67400 Illkirch, France*

\* These authors contributed equally.

† Deceased April 24th, 2018.

@ e-mail: bowen@unistra.fr

**Abstract**

Electrically manipulating the quantum properties of nano-objects, such as atoms or molecules, is typically done using scanning tunnelling microscopes[1–7] and lateral junctions[8–13]. The resulting nanotransport path is well established in these model devices. Societal applications require transposing this knowledge to nano-objects embedded within vertical solid-state junctions, which can advantageously harness spintronics[14] to address these quantum properties thanks to ferromagnetic electrodes and high-quality interfaces[15–17]. The challenge here is to ascertain the device's effective, buried nanotransport path[18], and to electrically involve these nano-objects in this path by shrinking the device area from the macro-[17,19–22] to the nano-scale[23–25] while maintaining high structural/chemical quality across the heterostructure. We've developed a low-tech, resist- and solvent-free technological process that can craft nanopillar devices from entire *in-situ* grown heterostructures, and use it to study magnetotransport between two Fe and Co ferromagnetic electrodes across a functional magnetic CoPc molecular layer[26,27]. We observe how spin-flip transport across CoPc molecular spin chains promotes a specific magnetoresistance effect, and alters the nanojunction's magnetism through spintronic anisotropy[28]. In the process, we identify three magnetic units along the effective nanotransport path thanks to a macrospin model of magnetotransport. Our work elegantly connects the until now loosely associated concepts of spin-flip spectroscopy[2,3], magnetic exchange bias[29,30] and magnetotransport[24,25] due to molecular spin chains, within a solid-state device. We notably measure a 5.9meV energy threshold for magnetic decoupling between the Fe layer's buried atoms and those in contact with the CoPc layer forming the so-called 'spinterface'[16]. This provides a first insight into the experimental energetics of this promising low-power information encoding unit[31].

**Main Text**

Recent research has unraveled how two electronic spins may interact with one another in the presence of an electric current between them. This interaction may be tuned thanks to the positional control of a scanning tunnelling microscope (STM)[1–3,7], but also through molecular design. For example, metal-organic phthalocyanines can form molecular columns such that superexchange interactions between the central atom of neighboring molecules can promote magnetic order. The nature of the metal atom and the molecular stacking geometry determines whether this order is ferromagnetic or antiferromagnetic and can exceed room temperature[26,27]. STM experiments have evidenced conductance jumps as a spectroscopic signature of electrically manipulating antiferromagnetic (AF) spin correlations along the chain (e.g. CoPc with spin S=1/2) from a ground state to well-characterized excited states[1].

However, in such experiments, setting a spin referential thanks to an external magnetic field[32,33] or using metallic ferromagnetic (FM) electrodes[34] is cumbersome. Lateral junctions[10,12] exhibit similar limitations, as well as interface quality issues[35], while lateral break junctions[8,9,11,13] can be tricky to form and are fragile. This has prevented the direct observation of the magnetoresistance due to this spin-flip current, or any insight into how this spin-flip current interacts with the magnetic properties of the FM/molecule interface.

Conversely, limited experiments on macroscale shadow-mask[17,19–22] and nanoindented[23–25] solid-state junctions that integrate FM electrodes and molecular layers reveal novel but so-far disjointed effects that conceptually overlap only loosely with STM-based reports, and with magnetometry at the FM/molecule interface. According to transport experiments, molecular adsorption onto a FM layer not only generates a highly spin-polarized interface[23] at room temperature[36] (nicknamed a 'spinterface'), but can attenuate the otherwise strong FM coupling between the FM layer's topmost and remaining monolayers[17]. According to magnetometry[29,30,37], this 'magnetic hardening' effect can complement the magnetic exchange coupling between a FM metal and AF-coupled molecular spin chains (MSCs). This effective magnetic exchange bias, and spin chain excitations, both appear to qualitatively alter magnetotransport[24,25], but the link has so far remained indirect and speculative.

Overall, this differing level of maturity between these model (STM, lateral) and solid-state junctions reflects not only a technological bottleneck toward the widespread study of high-quality vertical nanojunctions, but also a challenge to precisely determine the effective nanotransport path[18] within a solid-state device. In this Letter, we bridge these technological and knowledge gaps through spectroscopic temperature-dependent magnetotransport experiments. Herein, we craft solid-state vertical nanojunctions from full *in-situ* grown FM/CoPc/FM heterostructures using a low-tech nanojunction process that does not structurally degrade the active organic layer, or expose it to resist, solvents or air. We find that the magnetoresistance generated upon flipping the magnetic orientation of the two FM electrodes tracks the increase in junction conductance resulting from excitations along spin chains in the intervening CoPc molecular magnetic layer. According to a three-macrospin model that fits our magnetotransport data, the junction's FM electrodes are magnetically coupled to a central magnetic unit comprising MSCs coupled to the junction's lower interface. This strength of this unit's unidirectional anisotropy at 17K nearly doubles with spin excitations to effectively reach 18T at |V|>100mV, while its temperature dependence mimics that of exchange bias seen in magnetometry[29,30]. The magnetic coupling between this unit and the bottom FM layer abruptly drops with increasing T at ~70K, which pegs a 5.9+-0.3meV thermal energy threshold of the so-called 'magnetic hardening' effect[17] induced at the Fe/CoPc interface by molecule-induced charge transfer. We find that spin excitations dynamically induce a mutual magnetic coupling of all magnetic units, and alter their magnetic anisotropy as direct experimental evidence of spintronic anisotropy[28] due to this transport mechanism.

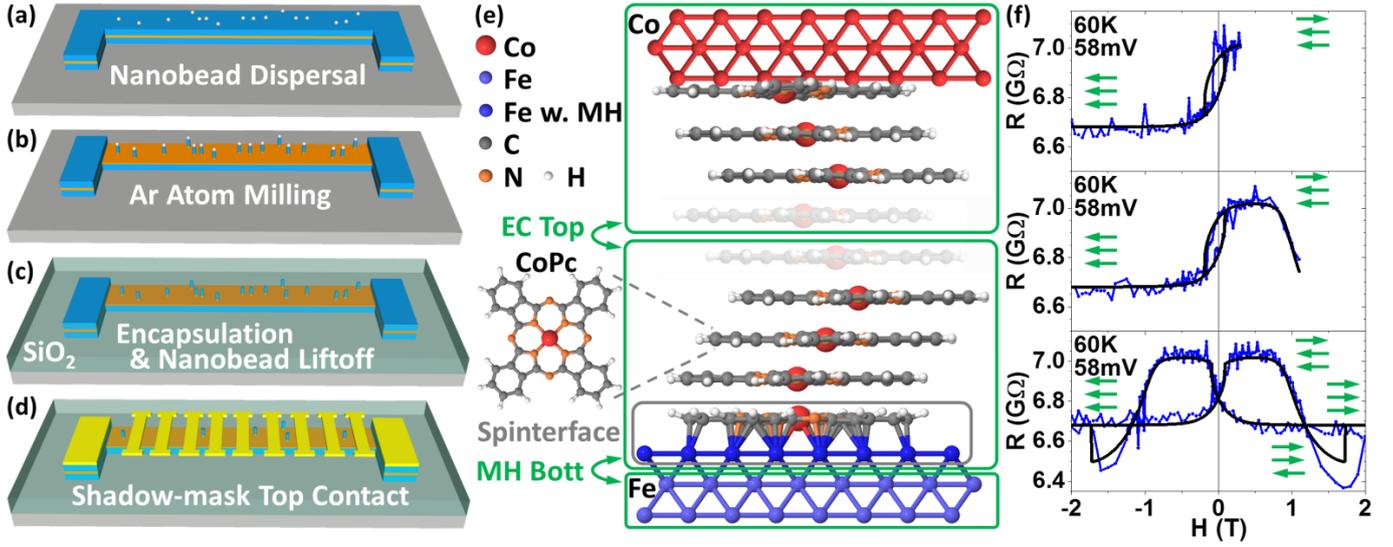

*Figure 1: Organic nanojunctions: processing and magnetics.* (a-d) Processing steps to transform entire heterostructure stacks into vertical nanopillars with dielectrically separated bottom and top contacts, without using resists/solvents. See text for details. (e) Schematic of the Fe/CoPc/Co junction's effective nanotransport path across three magnetic units (green boxes): a bottom Fe layer, the junction's lower interface forming a spinterface that is coupled to a MSC, and a top Co layer that is coupled to a MSC without a spinterface due to interdiffusion at the top interface. Not all molecular layers are shown (semi-transparent zone). The macrospin model's coupling terms ECTop and MHBott between the central and outer units are shown. See text for details. (f) Major and minor R(H) loops at T=60K and V=58mV reveal three R levels due to the magnetization reversal of the three magnetic units, which are schematized by green arrows. The fitting parameters were: $K_t/M_t$ =0.26T, $\theta_{Mt}$=60°, $K_c/M_c$ =4.95T, $\theta_{Mc}$=-4°, ECTop=-0.05T, MHBott=1.28T, $R_0$=6.68GΩ, MRTop=-2.4% and SpinFlipMR=5.1%.

Our junctions are crafted from entire *in-situ* grown FM/molecular layer/FM stacks, thereby preserving nominal structural/magnetic properties, especially at interfaces, using a novel, solvent-and resist-free processing technique inspired by nanosphere lithography[38] (see Methods for growth and processing details). After depositing the entire heterostructure through a shadow mask to define the lower electrode, 500nm-diameter $SiO_2$ nanobeads synthesized using a surfactant-mediated sol-gel reaction[39] are randomly distributed on the surface (Fig. 1a). Once Ar atom milling to the organic layer is complete (Fig. 1b), the sample is encapsulated in sputtered $SiO_2$ and the beads are blown off, leaving a sub-diameter access (not shown) to the nanopillar (Fig. 1c). In a final step, top metallic contacts are deposited through a shadow mask (Fig. 1d).

The nanotransport path[18] across the CoPc thin film nanojunctions that we infer from our combined experimental/analytical results is schematized in Fig. 1e. The effective nanotransport path proceeds across three coupled magnetic units. The energy density *E* of the nanotransport path is that of the top (i=t), central (i=c) and bottom (i=b) units and can be written as:

$$E = -\mu_0 \sum_{i=t,c,b} \boldsymbol{H}.\boldsymbol{M_i} + \frac{1}{2}\sum_{i=t,c,b} K_i \sin^2(\theta_{M_i} - \theta_{K_i}) - \sum_{i=t,b} C_{ci}\, \boldsymbol{m_c}.\boldsymbol{m_i}.$$

where $H$ is the applied magnetic field and, for each unit, $M_i$ is the magnetization, $m_i$ the reduced magnetization, $\theta_{M_i}$ the magnetization angle, and $K_i$ its uniaxial anisotropy with an easy axis angle $\theta_{K_i}$. Finally, $C_{ci}$ is the coupling strength between the central (c) and outer (i=b,t) magnetic units. We define $C_{ct}$ = *ECTop* and $C_{cb}$=*MHBott*. We justify this nomenclature hereafter. For each H step, *E* is minimized to yield, for each magnetic unit, the magnetization's in-plane orientation.

The adsorption of Pc molecules onto a FM surface induces a change in sign of spin polarization[15], as well as a magnetic hardening[40] for temperatures lower than ~70K[30] of the FM's top monolayer, which can rotate independently of the buried FM monolayers thanks to a weakened FM coupling that we call *MHBott*. The term 'spinterface' encompasses these interfacial properties, and sets the magnetic referential of the central magnetic unit, including that of MSCs that extend away from the interface and are formed by CoPc molecules with AF interactions[26]. Magnetometry confirms that these spin chains also contribute to the anisotropy below 100K[30]. The remainder of the bottom FM forms a lower magnetic unit that is modelled as a free layer, *i.e.* with a reduced anisotropy *K/M*=5mT.[1] As we will see, the top magnetic unit is comprised of the Co layer directly coupled to a MSC. The absence of a spinterface at the top interface presumably reflects metal interdiffusion during top electrode deposition. The presence in the nanotransport path of disjointed chains across the film thickness is expected from both structural and magnetic studies[26,30], such that a weak AF exchange term *ECTop* is present between the top and central magnetic units. The following dataset, which is typical of results found on several junctions (see Methods), was entirely acquired on a Fe/CoPc(20nm)/Co junction with resistances *R*(300K)=32kΩ and *R*(17K)=11GΩ at 20mV. This underscores thermally activated hopping transport[19,42] across the thin CoPc layer.

We first examine magnetotransport at 60K, i.e at the temperature onset of the magnetic hardening effect, and with MSCs already promoting exchange bias. Fig. 1 shows *R(H)* loops at 58mV following a cooldown at *H*=-1T. As the positive maximum *H* is increased, one notices a 1$^{st}$ resistance jump near *H*=0, and a 2$^{nd}$ resistance change centered around *H*=1.17T that is reversible as long as the resistance baseline at *H*=-2T isn't exceeded. For higher positive field sweeps, a third resistance change is observed and the *R(H)* loop becomes field-symmetric.

The three resistance levels observed justify our model's three magnetic units and suggest, consistently with further data/analysis, that two MR terms describe magnetotransport. We therefore write the resistance R due to non-collinear magnetizations $M_t$, $M_b$ and $M_c$ as:

$$R = R_0 \cdot \left[1 - \frac{MRTop}{2} \cdot (\mathbf{m_t} \cdot \mathbf{m_c} - 1) - \frac{SpinFlipMR}{2} \cdot (\mathbf{m_t} \cdot \mathbf{m_b} - 1)\right].$$

*MRTop* refers to the impact on nanojunction resistance of flipping the magnetization of the top magnetic unit relative to that of harder magnetic units, i.e. to the central magnetic unit. *SpinFlipMR* considers the MR due to flipping both the top and bottom magnetic units (see Suppl. Note 1 for details on pairing magnetic units to MR terms). We justify this nomenclature hereafter. By successfully fitting this unusual, symmetric *R(H)*, we identify the sequential magnetization reversal of the top, bottom and central magnetic units (see green arrows in Fig. 1f). Despite its low anisotropy, the bottom magnetic unit reverses at |*H*|>1T due to the FM coupling term *MHBott* arising from magnetic hardening (see Fig. 1e).

We experimentally define *MRTop*=*R*(2T)/*R*(-2T)-1 and *SpinFlipMR*= *R*($H_f$)/*R*(2T)-1 ($H_f$ = 1T for 17 < *T*(K) < 50, $H_f$< 1T for *T* > 50K, see Fig. 3), and schematize these MR contributions in the *R*(H) at 80mV and *T*=17K (see Fig. 2a), for which the magnetic states within |*H*|<2T are better defined. Due to our experimental limitation |H|<2T, when comparing the model with experiment, we make the approximation that $R_0 = R(-2T)$, while small deviations between experimental and modelled values of *MRTop* and *SpinFlipMR* can occur because full magnetization reversal of the bottom magnetic unit can be incomplete within |*H*|<2T. Nevertheless, dataset consistency criteria and the shape of the R(H) data strongly limit possible errors. See Suppl. Note 1 for details. The *ECTop* and *MHBott* coupling terms are also shown in Fig. 2a next to the reversal process for the top and bottom magnetic units, respectively.

---

[1] Only at the (V=80mV,17K) and (V=100mV, 72K) critical points does $K_b/M_b$ strongly increase, as expected in an exchange bias system near criticality[41]. See Suppl. Note 1 for details.

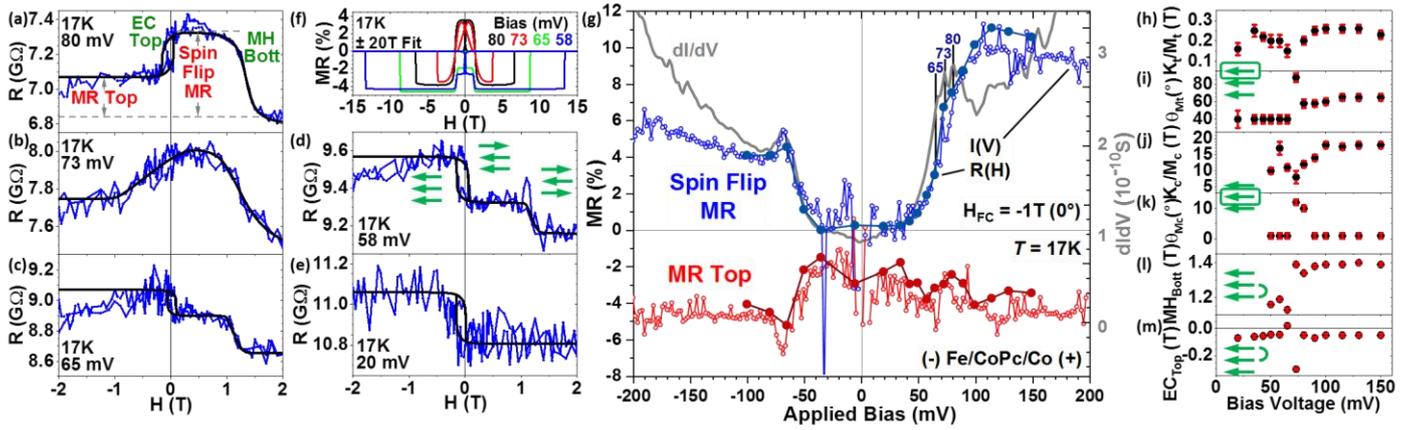

***Figure 2: Current-induced alterations to magnetotransport.*** *(a-e) R(H) loops at 17K for 20 < V(mV) < 80 and associated fits, also represented in panel f for ±20T. (g) Bias dependence of dI/dV, and of MRTop / SpinFlipMR inferred from I(V) and R(H) data. Bias dependencies of the strength and angle of the reduced anisotropy K/M for the (h-i) top and (j-k) central magnetic units. Bias dependencies of the (l) MH Bott and (m) EC Top coupling terms. The SpinFlipMR term tracks the dI/dV increase due to spin excitations. At the dI/dV peak, coupling between the MSCs of the top and central unit causes strong changes to their anisotropy parameters (panels h-k) and to the coupling parameters (panels l-m), which distorts the R(H) at 73mV (panel b). In the absence of the spin-flip conduction channel, flipping the bottom FM magnetization does not induce any MR (panel e) because the spinterface-stabilized MSC's AF ground state degrades spin-conserved transport. The error bars for data in panels h-m are discussed in Suppl. Note 1.*

We now examine magnetotransport at 17K. For 43 ≤ V(mV) ≤ 150, only two magnetization reversals are observed, as illustrated for R(H) loops in Fig. 2a-e. According to the model's fits (black lines of Fig. 2a-e), also shown in Fig. 2f for a larger field sweep, the central unit flips at H>2T, i.e. beyond our experimental limit. At 20mV, the R(H) loop (Fig. 2f) exhibits only one resistance change, even though the model's fit indicates that the bottom FM electrode has flipped during the sweep (data not shown). This implies that, at 20mV, *SpinFlipMR*=0. We will discuss this effect in what follows. Similar effects are observed for V<0 (data not shown).

Varying the in-plane orientation angle $\theta_H$ of the applied magnetic field causes *MRTop* to smoothly switch sign between 0º and 180º for *SpinFlipMR*=0 (see Suppl. Note 2). We refrain from studying *SpinFlipMR*($\theta_H$) since the magnetization state at H=1T evolves with $\theta_H$. Nevertheless, we observe that, after cooling from 120K at H=-1T, the R(H) at 17K and $\theta_H$ =0° is flipped along H when $\theta_H$ = 180° (see Suppl. Note 2). This flipping also occurs when comparing field cooling at H=-1T with field cooling at H=0⁺ after applying H=2T at T=120K (see Suppl. Note 2). Varying the amplitude of the bias voltage that is applied during cooldown does not significantly alter the R(H) loop (data not shown). We thus infer that the device's magnetic state during cooling defines a unidirectional axis for magnetotransport. As we will see, this unidirectional character arises from the coupling of the outer magnetic units to the central magnetic unit with very high anisotropy strength.

While magnetometry on FM/MPc bilayers revealed a fatigue effect upon repeatedly sweeping the external magnetic field[29,30], no such fatigue effect was seen in magnetotransport (see Fig. 1f). This, and the ability to reproduce magnetotransport data using a macrospin model (see Fig. 1f, Fig. 2a-e and Fig. 3a-e), suggest that the nanotransport path within the 500nm-nominal diameter is proceeding through a reduced number of grains.

Referring to Fig. 2g, the bias dependence of dI/dV reveals a mostly constant amplitude for |V|<35mV, and large increases for |V|>35mV, punctuated by peaks at |V|≈70mV. The dI/dV amplitude further increases at higher bias. In line with previous literature[1,2,25] on STM-assembled and solid-state-based transport across spin chains, we interpret these dI/dV features as the signature of spin excitations. We note that the complicated bias drop across the nominal 20 nm-thick CoPc junction due to the hopping transport regime could explain the higher bias onset compared to STM

studies[1] and impedes a discussion on the effective MSC length and exact spin excitations processes. This complicated bias drop, and the likely presence of magnetic disorder across the molecular layer (see hereafter), could also account for the different amplitudes in the *dI/dV* peaks for *V*>0 and *V*<0, instead of an interpretation as a signature of spin-polarized transport[2]. For this reason, we refrain from

We observe that *SpinFlipMR* spectroscopically tracks *dI/dV*, both from *R(H)* and *I(V)* data (see Fig. 2g). The small voltage lag originates from spectroscopic averaging effects for current compared to conductance[43]. To the best of our knowledge, this is the first observation of a MR signal that is driven to appear due to bias voltage, and whose amplitude tracks junction conductance. Given the above interpretation of *dI/dV*, this means that we have successfully measured MR between the two FM electrodes due to opening of spin-flip channels of transport across MSCs. The opposite signs of *MRTop* and *SpinFlipMR* are consistent with the change in sign of the spin polarization of the current that is expected[1–3,6,25] due to the spin-flip process across the MSC with S=1/2 spin moments. Indeed, the spin angular momentum thus conferred to the MSC causes the transport electron to flip its spin. This is then analyzed using the fixed spin referential of the collecting FM electrode in the two magnetization orientations. As a complementary effect, flipping the orientation of the injecting FM electrode's magnetization reverses the spin referential of the transport electron prior to a possible spin flip event. Thus, spin-flip transport channels effectively increase the junction conductance, but with an opposite sign of spin polarization, which is detected through the sign change between *MRTop* and *SpinFlipMR*.

The modelling of the bias-dependent *R(H)* data reveals that the reduced anisotropy strength $K_t/M_t$ of the top magnetic unit is ≈0.2T, *i.e.* over an order of magnitude stronger than that of a free layer (see Fig. 2h). We infer that it contains not only the top Co FM electrode, but also MSCs (see Fig. 1g). The weak, mostly bias-independent coupling term *ECTop* is then attributed to AF coupling between two MSCs belonging to the top two magnetic units along the nanotransport path.

With $K_c/M_c$≈10-20T, the central magnetic unit is magnetically very hard at 17K, and is coupled by *MHBott*~1T to the bottom magnetic unit. With a very low, almost always constant $K_b/M_b$=5mT, we infer that the bottom magnetic unit corresponds to the sub-interface atoms of the lower Fe electrode. *MHBott* describes the FM coupling expected between these layers and the spinterface owing to magnetic hardening. Comparing Figs. 2g and 2j, we see that the central unit's $K_c/M_c$ tracks the spin excitations. We thus infer that the central magnetic unit's spin referential is set by the spinterface and the MSC that is coupled to it. Our model indicates that, for *V*<35mV, the molecular layer's anisotropy must be at least ≈10T in order not to witness a symmetric *R(H)*.

Now that the nanotransport path's three magnetic units and couplings are identified and summarized in Fig. 1g, we examine how they are affected by spin excitations. While the *R(H)* loop for *V*=65mV (Fig. 2c) resembles that of *V*=80mV (Fig. 2a), reaching the *dI/dV* peak at 73mV results in a strongly distorted *R(H)* (Fig. 2b). To the best of our knowledge, this is the first evidence of how small bias changes can so drastically alter magnetotransport. This can be modelled only though a strong deviation in the angle of the central unit's anisotropy. We also witness strong deviations in the anisotropy strength/angle of the top and bottom magnetic units, while the AF coupling *ECTop* jumps by one order of magnitude. Meanwhile, the FM coupling *MHBott* jumps from 1.2T to 1.4T. We propose that spin excitations drive a dynamical magnetic coupling of the two otherwise disjointed MSCs. This would explain, for V>73mV, not only the increase in *MHBott* due to an effectively longer single MSC, but also how the top magnetic unit's anisotropy increases while its angle jumps and further increases in order to dynamically accommodate the geometrical requirements of the 90° super-exchange interaction between the two MSCs.

Overall, this strong impact on all three magnetic units at the bias voltage corresponding to maximum spin-flip conductance (see *dI/dV* peak at 73mV of Fig. 2g) can be interpreted as a mutual magnetic coupling between all three units that is induced by spin excitations. This constitutes direct evidence in magnetotransport of spintronic anisotropy, *i.e.* a change in magnetic anisotropy caused by a spin-polarized current[28,36], due to spin excitations.

In light of these observations, we now discuss the absence of any MR due to flipping the bottom FM layer for |V|≤35mV. Since *SpinFlipMR* does not abruptly reach a saturated value at the threshold bias voltage for the dI/dV increase, we conclude that the spin-flip transport channel is not altering the magnetic anisotropy of the bottom FM layer such that it can abruptly switch orientations. This is in line with the model's treatment of the bottom FM layer as a free layer subject to the strong FM coupling *MHBott* to the central magnetic unit's spinterface. We propose that, in the |V|≤35mV bias range, the MSC that is coupled to the spinterface impedes spin-polarized transport because it is in its AFM ground state and is structurally of high quality (high $K_c/M_c$). In that case, only the spin flip channel can reveal MR due to flipping the bottom FM layer's magnetization. We presume that this doesn't occur at the top interface because of structural imperfections that are embodied by the top magnetic unit's much lower $K_t/M_t$.

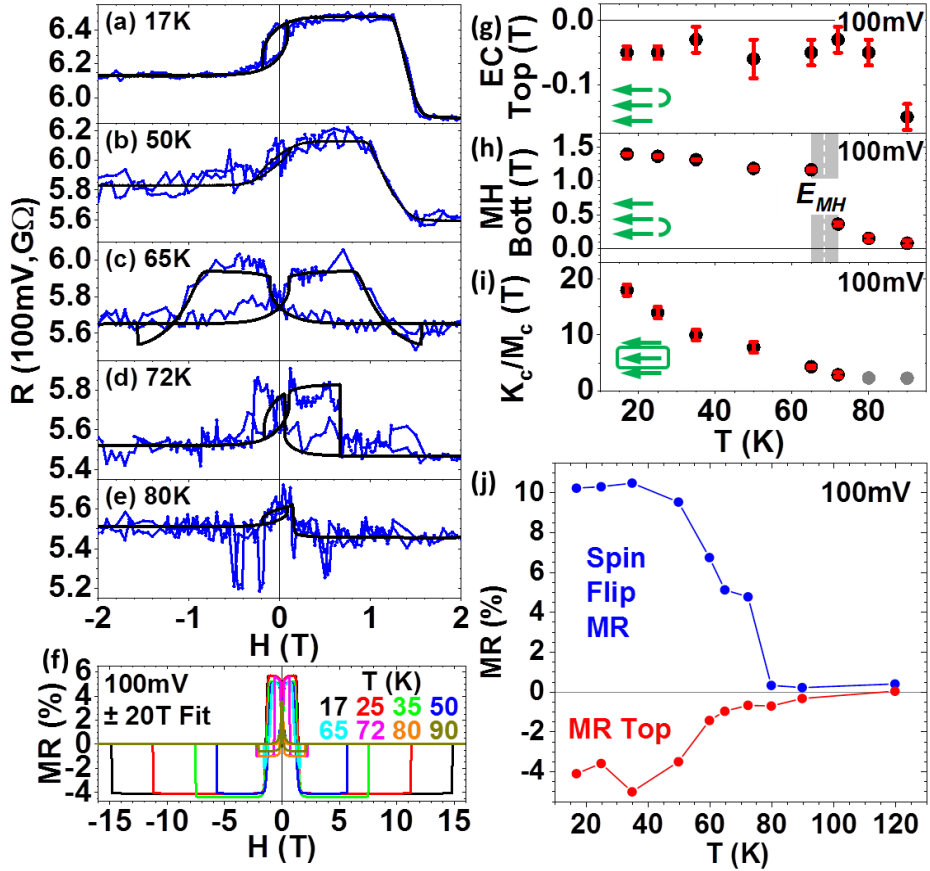

*Figure 3: Temperature weakens spin chain coupling and magnetic hardening.* (a-e) R(H) loops at 100mV for 17 < T(K) < 80. Data are in blue, while the modelled fits are in black and are also shown in panel f for ±20T. Temperature dependencies of (g) ECTop, (h) MHBott, (i) $K_c/M_c$ of the central magnetic unit and (j) of the SpinFlipMR and MRTop. The gray data in panel i for T≥80K are the minimum values required to obtain R(-2T)≠R(2T). The error bars for data in panels g-i are discussed in Suppl. Note 1.

Increasing temperature causes the nanojunction magnetics, and thus magnetotransport, to strongly change, as illustrated by the *R(H)* loops at 100mV for 17 < *T(K)* < 80 of Fig. 3a-e. The increasing noise despite the *R* decrease suggests that thermal excitations are destabilizing magnetism along the nanotransport path. For 17 < T(K) < 55, the same flipping of the top and bottom magnetic units is observed for |H|≤2T (data at 55K not shown). For 60 ≤ *T(K)* ≤ 72, a symmetric *R(H)* is observed (see also Fig. 1f). For *T*≥80K, the *R(H)* loop has collapsed, and only low-field MR is observed until 100K. Thus, only a 5-8K temperature increase around 55K and 75K can promote remarkable changes in the *R(H)* loop shape.

Modelling these *R(H)* loops (black lines of Fig. 3a-e and Fig. 3f) reveals that, while *ECTop* remains mostly constant, weak and negative (i.e. AFM coupling, see Fig. 3g), *MHBott* decreases only moderately up to 65K, and then abruptly decreases for T>65K (Fig. 3h). We thus peg a $E_{MH}$=5.9±0.3meV threshold energy for the magnetic hardening effect at the *bcc* Fe(110)/CoPc interface, which constitutes a first experimental energetic benchmark toward encoding information using the spinterface[31]. For thermal fluctuations above this energy threshold, the spinterface and bottom FM electrode can no longer rotate independently, which explains the collapse for *T*>72K of the *R(H)* loop (Fig. 4e).

According to the model, the central unit's anisotropy $K_c/M_c$ decreases steadily from ≈18T at 17K to ≈4.4T at 65K, and then to under 3T for *T*>70K (Fig. 4i). This anisotropy term is thus dominated by a MSC contribution, but also comprises a contribution due to magnetic hardening, in line with the model's description of the molecular layer (see Fig. 1a). Both this term and *MHBott*, both cause *SpinFlipMR* and *MRTop* to concurrently decrease with increasing *T*, and to disappear at *T*=100K, thereby mimicking the effective exchange bias effect seen in magnetometry data[29,30].

To conclude, our work articulates the concepts of spin-flip spectroscopy, magnetic hardening and exchange bias at ferromagnetic metal/molecule interfaces (so-called 'spinterfaces'), and spintronic anisotropy within magnetotransport across solid-state nanojunction devices. By using FM electrodes with a fixed spin referential for transport, we isolated the magnetoresistance contribution arising from spin flip excitations along molecular spin chains. In the process, we identified three coupled magnetic units along the spintronic nanotransport path, and examined how bias voltage and temperature drive the nanopath's magnetism, thanks to a phenomenological macrospin model of transport. This also showed how spin excitations alter the nanopath's spintronic anisotropy. Temperature-dependent studies reveal the energy threshold for magnetic decoupling of the spinterface, and the complementary impact on magnetotransport of both the MSC and of the magnetic hardening effect, beyond magnetometry data on FM/molecule bilayers[29,30]. This establishes an experimental benchmark into the energetics of encoding information using the spinterface[31]. This nanotechnological progress is enabled by an innovative, low-tech, solvent- and resist-free processing technique that works with entire heterostructure stacks and can be further rationalized (e.g. using nanobead positioning techniques) toward industrial applications of quantum physics using nano-objects within solid-state devices.

**Methods**

Heterostructure stacks were grown in-situ and at room temperature in an ultra-high vacuum multichamber cluster by dc sputtering (metals) and thermal evaporation (CoPc). The $SiO_x$ substrate was annealed at 110°C and allowed to cool down prior to deposition. After nanojunction processing (see main text), the junctions were wirebonded to a sample chip and inserted onto a cryo-free magnetotransport bench. Measurements were performed in 4-point mode with (-) contacts on the lower electrode. In the main text, the junction stack was $SiO_x$//Cr(5)/Fe(50)/CoPc(20)/Co(10)/Cr(5) (all numbers in nm). In addition to that junction, eight CoPc junctions with both top and bottom Fe electrodes showed MR for *T*<100K, fived exhibited spin-flip behavior, and MR tracking of junction conductance was observed on two junctions.

**Acknowledgements**

We thank Y. Henry for stimulating discussions, M. Gruber for reading our manuscript, S. Dornier for engineering assistance, the STNano technological platform staff for technical assistance with certain processing steps, and the remaining members of the IPCMS machine shop for support. We acknowledge financial support from the Institut Carnot MICA (project 'Spinterface'), from the Region Grand Est and Synchrotron SOLEIL, from CEFIPRA grant 5604-3, from the ANR (ANR-06-NANO-033-01, ANR-09-JCJC-0137, ANR-14-CE26-0009-01), the Labex NIE "Symmix" (ANR-11-LABX-0058 NIE), the EC Sixth Framework Program (NMP3-CT-2006-033370), the Contrat de Plan Etat-Region grants in 2006 and 2008, by the Region Grand Est 'NanoteraH' project supported by the EU through the European Regional Development Fund program, by the impact project LUE-N4S part of the French PIA project "Lorraine Université d'Excellence", reference ANR-15IDEX-04-LUE and by the « FEDER-FSE Lorraine et Massif Vosges 2014-2020 ", a European Union Program.

**Author Contributions**

E.B., M.B. and S.B. conceived the experiments. E.U., F.S and B.C. grew samples with assistance from J.A., V.d.C. and C.K.. E.U. developed the nanojunction process. E.U. and L.K. crafted the nanojunction stacks with assistance from D.M., K.K., A.B., F.S. and J.A.. M.B., U.H., B.L., N.B., D.S., P.P., A.B. and E.S. implemented the magnetotransport infrastructure. K.K., M.B., B.C. and L.K. measured MTJs. K.K. and M.B analyzed the data with input from S.B., E.B., W.W. and M.A.. D.L., M.H. and F.M. proposed the 3-macrospin model approach. D.L. implemented the magnetotransport model with input from M.B. . M.B., K.K., D.L. and M.H. fitted the experimental data. M.B. wrote the manuscript with input from S.B., K.K., W.W., F.N. and D.L..

*Magnetoresistance and spintronic anisotropy induced by spin excitations along molecular spin chains*

*Supplementary Information*

K. Katcko[1]*, E. Urbain[1]*, L. Kandpal[1], B. Chowrira[1,2], F. Schleicher[1], U. Halisdemir[1], F. Ngassamnyakam[1], D. Mertz[1], B. Leconte[1], N. Beyer[1], D. Spor[1], P. Panissod[1], A. Boulard[1], J. Arabski[1], C. Kieber[1], E. Sternitsky[1], V. Da Costa[1], M. Alouani[1], M. Hehn[3], F. Montaigne[3], A. Bahouka[4], W. Weber[1], E. Beaurepaire[1†], D. Lacour[3], S. Boukari[1], M. Bowen[1@]

**Supplementary Note 1 : Fitting R(H) data using the macrospin 3-unit model**

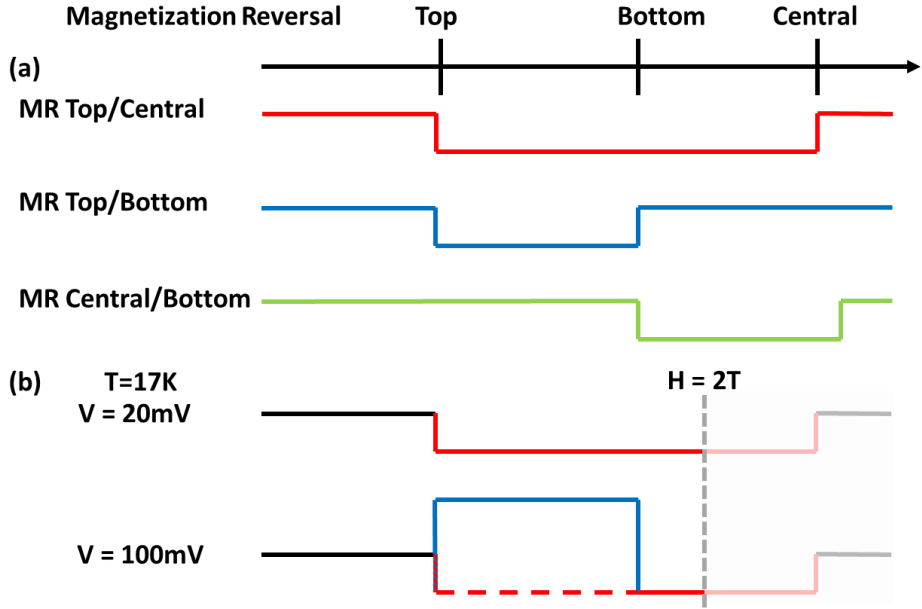

*Suppl. Fig. S1: Rationalization of magnetization reversals and magnetoresistance (MR) terms.* (a) MR traces upon flipping pairs of magnetic units. MR<0 is assumed in all cases for simplicity. (b) The experimental MR data for T=17K and V=20mV is most easily fitted using one MR Top/ Central term. Since the R change upon flipping the top unit magnetization switches from a decrease to an increase as V is increased, the MR trace at V=100mV is most easily fitted by combining two MR Top/Central and MR Top/Bottom terms. Other combinations are possible but not straightforward as they would require the perfect cancellation of MR terms. See text for details.

This Note details the procedure that was used in order to fit the *R(H)* data at 17K versus applied bias voltage, and at 100mV versus temperature. The macrospin 3-unit model contains several parameters : two for each magnetic unit (Top unit=t, Central unit=m, Bottom unit=b; anisotropy strength *K/M* and angle θ), the two coupling terms *ECTop* and *MHBott* between the central and outer magnetic units, and two MR coefficients. *MRTop* is associated with flipping the magnetization of the top and central layers (called 'MR Top/Central in what follows), and *SpinFlipMR* with that of the top and bottom layers (called 'MR Top/Bottom' in what follows). This association proceeds from the observation of the *R(H)* curves at 17K (main Fig. 2) and 60K (main Fig. 1) under the assumption of the sequential magnetization -reversal of the three magnetic units. Indeed, referring to Fig. S1, three MR terms may potentially define the effective *R(H)* data. Each MR term has a specific H-dependent signature (see Fig. S1a). We are able to fit our data using only *MR Top/Central* and *MR Top/Bottom* terms. As schematized in Fig. S1b, the (17K, 20mV) data can be fitted using only a *MR Top/Central* trace (compare with Fig. 2e). Furthermore, we experimentally observe that the MR upon reversing the top unit is made to eventually switch sign as the spin excitation conductance channel is opened with increasing bias. This is also associated with a MR term upon flipping the bottom magnetic unit's magnetization. These two aspects are schematized in Fig. S1b for the case of (17K, 100mV; see Fig. 2a for a similar experimental trace at 80mV): the R(H) data can be reproduced by adding to the *MR Top/Central* trace a *MR Top/Bottom* trace.

Within the self-consistent description of the three magnetic units provided in the main text (see Fig. 1e), we constrained the fitting procedure by imposing 'free-layer' properties to the bottom magnetic unit: $K_b/M_b$=5mT. The external magnetic field at which the magnetization of the bottom magnetic unit reverses is then driven mostly by its coupling *MHBott* to the Central magnetic unit (see Fig. 2l). The shape of the reversal (sharpness of the onsets, slope of the reversal) is determined by the central layer's anisotropy strength $K_c/M_c$, while $\theta_m$ was almost always constant (see Fig. 2j-k). The top magnetic unit's $K_t/M_t$ and $\theta_t$ were fitted in order to reproduce the minor loop due to the two magnetization reversals at low *H*. Here, care during fitting was taken to adhere to the 'squareness' of the loop, and to the evolution from saturation to remanence. The loop's bias shift was set through *ECTop*. This ability to compartmentalize the fitting parameters to segments of the *R(H)* trace constitutes an important measure of the confidence of the fitting procedure. In panels h-m of Fig. 2, the error bars show, with all other parameters constant, the parameter range within which the fit is still considered correct after visual inspection. We present in Suppl. Fig. S2 examples of the R(H,17K,115mV) data to showcase this aspect.

This compartmentalization of our 3-macrospin model's parameters to *R(H)* features fails at the critical points (17K, 73mV; point 'A'; see Fig. 2b) and (17K, 80mV, point 'B'; see Fig. 2a) (i.e. on the dI/dV peak, see Fig. 2g), and at (72K ,100mV, point 'C'; see Fig. 3d) (i.e. when the magnetic hardening is overcome by thermal fluctuations). At points 'B' and 'C', it is impossible to maintain 'free-layer' properties for the bottom magnetic unit: $K_b/M_b$~3000 and $\theta_b$=13°. At points 'A' and 'B', $\theta_m$≈11°. Note also the deviations in *ECTop* and $\theta_t$ at point 'A', which are associated with a jump in *MHBott* to a higher, nearly constant value. These mutual couplings of otherwise independent model parameters indicate that the three magnetic units become dynamically coupled at these critical points due to changes induced by the spin-polarized current in the spin-flip conductance channel's fully spin-polarized current, i.e. by spintronic anisotropy[2] (points 'A' and 'B') and by thermal fluctuations (point 'C').

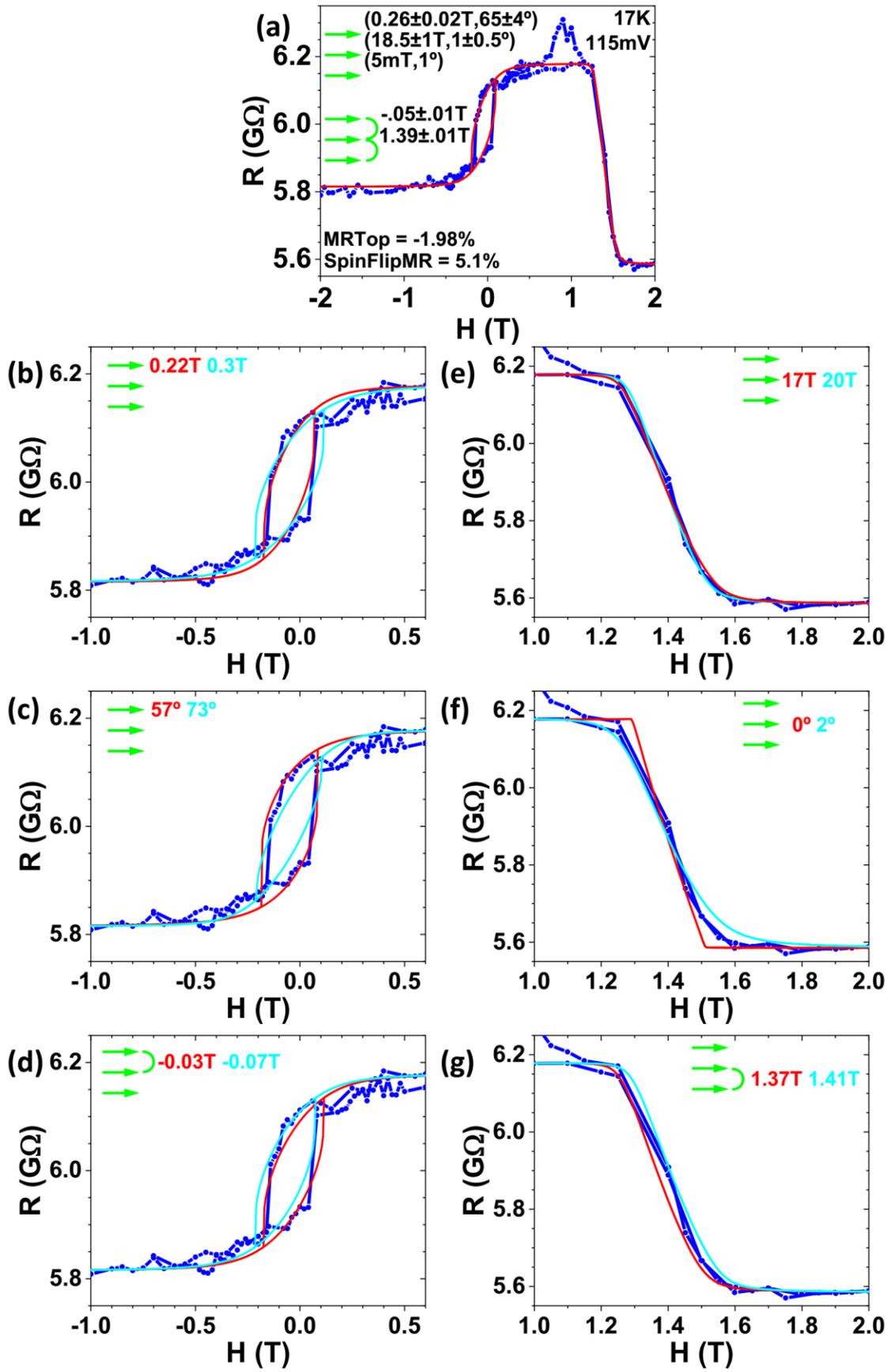

**Suppl. Fig. S2: Determining the error bars of the fit.** (a) The data and final fit for R(H) at 17K and 115mV. Parameters with error bars are shown. Zooms on experimental data and two fits outside the error margin for (b) $K_t/M_t$, (c) $\theta_t$, (e) ECTop, (e) $K_t/M_t$, (f) $\theta_c$ and (g)MHBott.

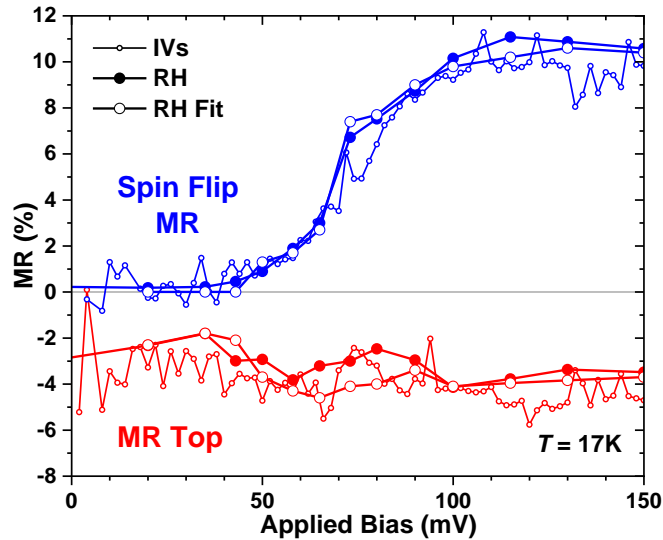

***Suppl. Fig. S3: Experimental and fitted MR terms.*** *Experimental MRTop and SpinFlipMR obtained from IV and R(H) data, and fitted MRTop and SpinFlipMR parameters.*

To fit the R(H) data, the two *MRTop* and *SpinFlipMR* terms are used. However, the *R(H)* traces do not necessarily guarantee full magnetization reversal within the |H|<2T experimental window, if at all. Fitting therefore proceeded starting with high bias voltage data (see e.g. (17K, 115mV) data in Suppl. Fig. S2), for which the resistance is mostly flat not only at large H<0, but also for 1.7 < H(T) < 2, under the assumption that these plateaus imply full magnetization reversal. As V or T was changed, the evolution of the plateau at 1.7 < H < 2 was monitored, and the *MRTop* and *SpinFlipMR* values adjusted accordingly to mimic the R(H) slope in this H range. Suppl. Fig. S3 presents the experimental and fitted MR values found for V>0. Despite small deviations, the fitted MR terms tracks their experimental counterparts rather well. This supports the model's initial hypotheses relating to the initial fully aligned magnetic state at H=-2T, and the definitions of *MRTop* and *SpinFlipMR* in the main text.

**Supplementary Note 2 : Magnetic anisotropies along the nanotransport path**

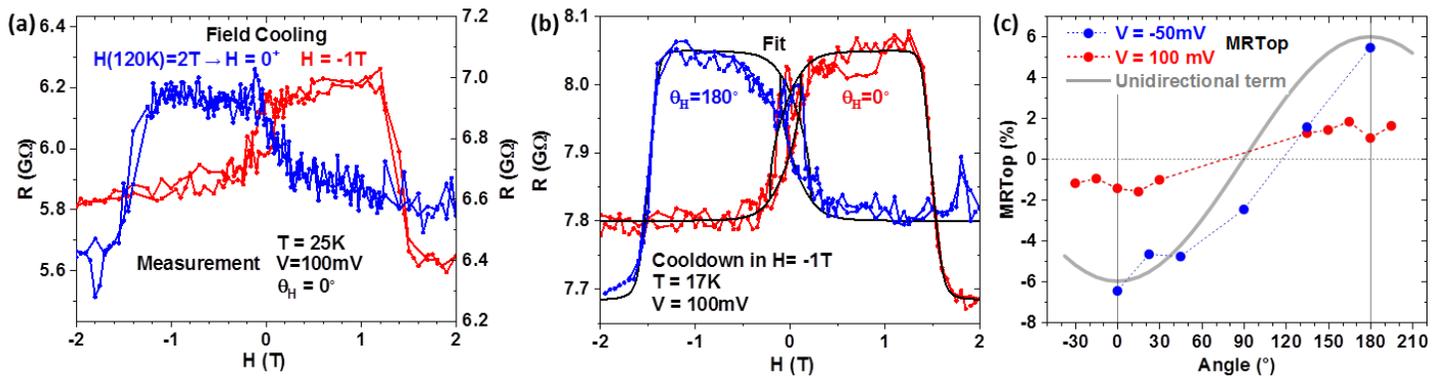

***Suppl. Fig. S4: Determining magnetic anisotropy from magnetotransport.*** *(a) R(H) acquired at 17K and 100mV after field cooling in H=-1T (red) and H=0⁺ T (blue; H=2T was first applied at 120K). Field-cooling at H<0 defines the in-plane angle θ=0. (b) R(H) acquired at 17K and 100mV for $\theta_H$=0° (red) and 180° (blue). After fitting the $\theta_H$=0° data, the same set of parameters was used, but the simulation was run for $\theta_H$=180°. (c) Angular dependence of MRTop at 17K for V=-50mV (i.e. without spin excitations along the MSCs) and for V=100mV (i.e. with spin excitations along the MSCs). Variations in*

*R(H=-2T) between panels reflects a minor evolution of junction resistance during the ~100-hour measurement and across several field-cooling sequences. The fitting parameters in panel b were: $K_t/M_t$ =0.28T, $\theta_{Mt}$=70°, $K_c/M_c$ =21T, $\theta_{Mc}$=-1°, ECTop=-0.05T, MHBott=1.49T, $R_0$=7.8GΩ, MRTop=-1.4% and SpinFlipMR=4.8%. Small differences between these parameters and those found in the main text reflect minor evolutions in junction magnetotransport during the measurement run.*

To test the conditions for, and the symmetry of, magnetic anisotropies along the nanotransport path, we first examined cooldown conditions. We present in Suppl. Fig. S4a R(H) data at 25K and 100mV mV after cooling the junction from 120K while maintaining $\theta_H$=0. The red R(H) data is obtained after cooling in H=-1T. If, after applying H=2T at 120K, cooldown is instead performed at H=0+, this causes the *R(H)* trace to flip about the *H*=0 axis. This also occurs if, while at 17K, the in-plane angle of the external magnetic field *H*, $\theta_H$, is switched from 0° to 180° (Suppl. Fig. S4b). Cooling down with *H*=1T applied at an in-plane angle of 90° instead of 0° did not change the *R(H)* loop (not shown). The angular dependence of *MRTop*, shown in Fig. S4c, reveals a unidirectional behavior both without (V=-50mV; see Fig. 2g)) and with (V=100mV) the spin-flip conductance contribution. Due to the additional *SpinFlipMR* term in the latter case, and given our |H|<2T measurement window, it isn't possible to follow the angular dependence of *MRTop* for 30 < $\theta_H$ < 135 because there is no longer a *R* plateau for 1.7 < *H*(T) < 2. Note that this, and the fact that the 1T resistance level becomes undetermined at intermediate angles, are why it is not possible to study the angular dependence of *SpinFlipMR*. Generally, we conclude from these data that it is the orientation of magnetization of the FM electrodes during cooldown that determines a unidirectional anisotropy along the nanotransport path.

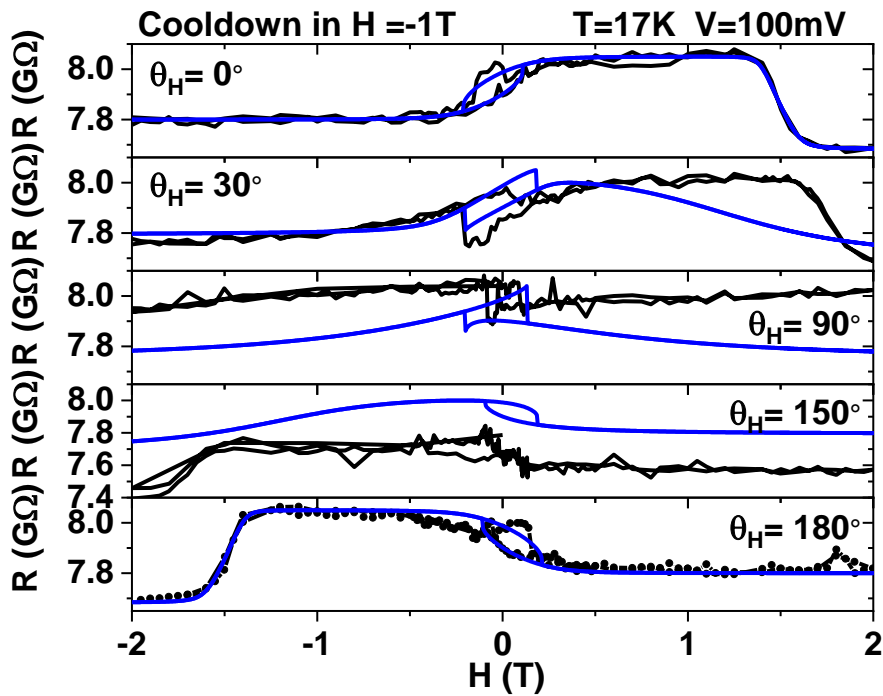

**Suppl. Fig. S5: Angle-dependent R(H): experimental & fitted data.** *Parameters are the same as those in Suppl. Fig. S4b.*

To be complete, we show in Fig. S5 experimental and fitted *R(H)* data at (17K,100mV) for salient values of the in-plane angle $\theta_H$ of the applied magnetic field. For deviations from the field-cooling angle $\theta_H$=0, the fitted R(H) obtained by simply adjusting the in-plane angle of H while conserving the same parameters only qualitatively reproduce features of the experimental *R(H)*, whereas an exact fit is obtained when the fitting parameters found for $\theta_H$=0 ° are used for $\theta_H$=180° (see also Suppl. Fig. S4b). This might reflect limitations of our model's macrospin assumption, or the presence of higher-order terms within a more complicated model. Nevertheless, our model captures all essential magnetotransport features (see Figures of main text).